\documentstyle[cjaa209,twoside,epsf]{article}
\input{psfig.sty}                               

\begin{document}

\title{Asteroid Deflection: How, where and when? }

\author{D. Fargion $^{1,2}$
\mailto{daniele.fargion@roma1.infn.it}
}

\inst{
$^1$ Physics Department.Rome Univ.1, Sapienza, Ple A.Moro 2, 00185, Rome,Italy\\
$^2$ INFN Rome Univ.1, Italy\\
}

\email{daniele.fargion@roma1.infn.it}

\markboth{Daniele Fargion}
{Asteroid Deflection: How, where and when? }

\pagestyle{myheadings}
\date{Received~~2007~~~~~~~~~~~~~~~ ; accepted~~2007~~~~~~~~~~~~~~ }

\baselineskip=12pt

\begin{abstract}
\baselineskip=12pt To deflect impact-trajectory of massive  and
spinning $km^3$ asteroid by a few terrestrial radiuses one need a
large momentum exchange. The dragging of huge spinning bodies in
space by external engine seems difficult or impossible. Our solution
is based on the landing of multi screw-rockets, powered by
mini-nuclear engines, on the body, that dig a small fraction of the
soil surface to use as an exhaust propeller, ejecting it vertically
in phase among themselves. Such a mass ejection increases the
momentum exchange, their number redundancy guarantees the stability
of the system. The slow landing (below $ \simeq 40 cm s^{-1}$) of
each engine-unity at those very low gravity field, may be achieved
by safe rolling and  bouncing  along the surface. The engine array
tuned activity, overcomes the asteroid angular velocity. Coherent
turning of the jet heads increases the deflection efficiency. A
procession along its surface may compensate at best the asteroid
spin. A small skin-mass (about  $2\cdot10^4$ tons) may be ejected by
mini-nuclear engines. Such prototypes may also build first safe
galleries for humans on the Moon. Conclusive deflecting tests might
be performed on remote asteroids. The incoming asteroid 99942
Apophis (just 2$\%$ of $km^3$) may be deflected safely a few Earth
radiuses. Its encounter maybe not just a hazard but an opportunity,
learning how to land, to dig, to build and also to nest safe human
station inside. Asteroids amplified deflections by gravity swing may
be driven into longest planetary journeys, beginning i.e. with the
preliminary landing of future missions on Mars' moon-asteroid Phobos
or Deimos.

\end{abstract}

\section{Introduction: the asteroid safe deflection}
On late December 2004 an asteroid labeled 2004 MN4, now named 99942
Apophis, has been noted  as  being in possible future collision
trajectory with Earth on 2029 and 2037. The 2029 nearer encounter
will be as near as 5 terrestrial radiuses. Therefore we shall
inquire the need to divert in a decade time by the same distance an
incoming spinning  asteroid.

The case is now less pressing because a better trajectory knowledge
reduced the future impact probability. Nevertheless the subject
revived  the simple question on how and when it is possible to
deflect an asteroid. The possibility of an asteroid collision with
Earth is  usually small but, as in the past, in any near or far
future it may be faced as a real threat to human life or in general
to life. Here we re-derived the simplest basic law ruling the
asteroid deflection, keeping in mind its very possible spinning
nature, in a quite general way.

The landing of a screw-rockets array on a spinning asteroid to dig
the soil, to fuel engine and to deflect its trajectory in phase,
possibly while contra-rotating, is the proposal of the present
paper.

In next Section 2 we show in parametric form the  general deflection
laws, linking energy budget, asteroid mass, jet velocity and mass to
eject in a unique system. In next Section 3 we discuss   why mass
propulsion is much better than  radiation pressure, taking into
account the prompt nuclear options. In the main Section 4 we compare
the spin energy versus the deflection one, showing the difficulty to
deal with the asteroid angular momentum. We also consider the
landing of the screw-engine unities and their cooperative deflecting
role on the whole asteroid surface by three main processes: Array in
phase, Turning Jets, Contra-rotating procession. In Section 5 we
conclude  that deflecting asteroids by array is possible: this
effect can be amplified by gravity swing. These large deviation may
drive them in space, offering  a safe  housing, from cosmic ray
radiation flare, for human travels in far planetary spaces. Some
applications to near future 99942 Apophis event are considered in
conclusions section, where we present a surprising note on the key
role of Phobos and Deimos (being asteroid-moons of Mars): these
moons are in fact ideal guest place before Mars landing.

In late appendix (A,B,C,D,E,F) we answer to six  related questions:
How many Nukes for a deflection? Why Neutron Bombs are better in
impulse transfer? May Asteroid be charged and deflected  by Solar
Lorentz Forces? Are hypothetical Anti-Asteroids deflections, by gas
annihilation, observable? May fast Spinning  Asteroid be break down
and split in pairs? The last question is  puzzling also considering
the relevant spinning effect made by tidal forces while Apophis will
approach at a few radius impact distance the Earth on 2029
encounter. The final and actual question is how to tag successfully
Aphofis trajectory on 2017.

\section{The basic parametric laws}
We shall first summarize the equation of motion for an asteroid
deflection. In the large mass limit $M_A \gg m $ of the asteroid
$M_A$ respect the ejected fuel one $m$ and in a non relativistic
approximation for propeller velocity ($v_1 \ll c$), the main
equation of motion are ruled by the needed deviation distance d
(let's scale it by $5\cdot R_{\oplus}$), the mass of the flying
object $M_A$ (let's scale it by a $km^3$ water like mass or giga-ton
unity) in the given time $t$ that may be considered in a
characteristic decade duration length, and also by a given total
engine energy  budget $E_t$ (we compare it also with Hiroshima bomb
energy unit $\simeq 10^{21}$ erg).  The consequent needed asteroid
acceleration ($a_{Asteroid}$) is:

\begin{equation} {a_{Asteroid}}=\frac{2d}{t^2}=6.4\cdot10^{-8}\,cm s^{-2}(\frac{d}{5\cdot R_{\oplus}})(\frac{t}{10
y})^{-2} \label{1}
\end{equation}

The required asteroid final velocity is:

\begin{equation}
{v_f}=\frac{2d}{t}=20.2\,cm s^{-1}(\frac{d}{5\cdot
R_{\oplus}})(\frac{t}{10y})^{-1} \label{2}
\end{equation}

The consequent required thrust  (force) is:

$$
{F}_{Asteroid}=M\cdot a_{Asteroid}={\dot{m}}_{ejected}\cdot
v_1=2Md\cdot t^{-2}=$$
\begin{equation}
{F}_{Asteroid}=640.5N(\frac{M}{10^{15} g})\cdot(\frac{d}{5\cdot
R_\oplus})\cdot(\frac{t}{10 y})^{-2} \label{3}
\end{equation}

This small force is comparable with a terrestrial person weight.
Just to compare the above force with the one due to the  radiation
pressure of solar light at 1 AU over a rock sphere whose mass does
correspond to a $km^3$ water one with a typical radius of
$R=4.51\cdot10^4$ cm, and whose disk area is $A=0.64\,km^2$ or more
simply on an ice-like sphere asteroid whose radius is now
$R=6.2\cdot10^4$ cm, and whose disk area is $A=1.2\,km^2$, one finds
for the latter $4.66$ or $9.3$ Newton, respectively for total
absorption or complete retro-reflection. A possible large
mirror-sail may mildly deflect the body asteroid, but the needed
deflection force considered above (eq.\ref{3}) is at least two or
three order of magnitude above the radiation one, without any
consideration on the problematic about the safe holding of a sail on
a spinning body. It is nevertheless worth full to consider the
possible mild deflection (hundreds km size in a decade) achieved by
a complete albedo change. There are at least three very different
changes: \emph{black spray painting}, \emph{mirror layer painting}
or \emph{retro-reflector painting}. The total  paint mass may reach
a few tons (for few micron thickness). The eventual painting
deflection, mass and operative cost maybe comparable or competitive
with the Kinetic Impactor deflection \cite{Rizzo et all 2006}, but
the final deflection do not reach the needed few Earth radiuses
desired. Most the final kinetic energy $E_{k}$, (of the total
initial energy $E_t$ stored in the rocket engines, partially
dissipated in heat) resides in the fast exhausted gas mass (m),
$E_{k1}$, ejected at  velocity ($v_1$), while only a tiny  fraction
of the total kinetic energy lay in the slow asteroid, at final
velocity $v_f$, whose final deflected momentum is $\Delta P_{
Ast}=2.018\cdot 10^{16} g\cdot cm s^{-1}(\frac{d}{5\cdot
R_{\oplus}})(\frac{M}{10^{15} g})(\frac{t}{10y})^{-1}$

\begin{equation}
E_{k Ast}=2.035\cdot10^{17}erg\cdot(\frac{d}{5\cdot R_{\oplus}})^2(\frac{t}{10y})^{-2} (\frac{M}{{10^{15}g}})
\end{equation}

To be more precise the exact final velocity $v_{Asteroid-final}$
derived by equation of motion for a rocket at a final mass $M_f$ and
initial one $M_i$ ($M_f - M_i=m$) is: $v_{Asteroid-final}=v_1
\cdot\ln[\frac{M_i}{M_f}]=v_1\cdot\ln[\frac{M_f + m}{M_f}]$, where
$v_1$ is the emission velocity of the propeller mass $m$ respect to
the asteroid. However because $M_f\gg m$ the above linear
approximation holds.

In general $\frac{E_{k Ast}}{E_{k1}}=\frac{m}{M}$; for present
characteristic $km^3$ events $\frac{E_{k
Ast}}{E_{k1}}=2\cdot10^{-5}(\frac{m}{2 \cdot 10^{10}
g})(\frac{M}{10^{15} g})^{-1}$. Because of momentum conservation,
for any given engine efficiency $\eta$ in converting into kinetic
thrust the total rockets energy $E_t$, we have $\eta\equiv
\frac{E_k}{E_t}\leq1$ and the needed engine-jet velocity $v_1$ for
any asteroid mass, deflection time, displacement and total energy,
are bounded at once in the expression:

\begin{equation} {v_1}=\frac{E_t\cdot\eta\cdot t}
{M\cdot d}=9.9\cdot10^5 cm
s^{-1}\eta\cdot(\frac{E_t}{10^{22}erg})(\frac{t}{10
y})(\frac{M}{{10^{15} g}})^{-1}(\frac{d}{{10 y}})^{-1}
\end{equation}

Here we imagined a characteristic reserve energy for all the engine,
of the order of $E_t=10^{22}erg=10^{15}joule$ unity, corresponding
to nearly ten Hiroshima bomb energy which is a typical value for
small size nuclear engines. The nuclear fuel engine, spread in a
dozen of unities, should not exceed a few tenths of kilos each one
while the nuclear core, the robotic screw engines, the tractor
wheels, the thruster rockets, might be well within a half a ton mass
per unit. We remind the absence of any radiative protection in these
robotic unities, allowing a light unit reactor mass. The smaller and
the lighter, the better. The relation between the exhaust soil
kinetic and total energy conversion $E_{k_1}=1/2\cdot m
v_1^2=E_t\cdot\eta=10^{22}erg\cdot\eta$ and the above equation,
defines the needed total propelled mass:

\begin{equation}
{m}_{ejected}=\frac{2\cdot M^2 d^2}{E_t \cdot\eta\cdot
t^{2}}=2.037\cdot10^{10}\,g\;(\frac{M}{10^{15} g})^2(\frac{d}{5\cdot
R_\oplus})^2(\frac{E_t}{10^{22}\,erg})^{-1}(\frac{\eta}{1})^{-1}(\frac{t}{10
y})^{-2}
\end{equation}
\begin{equation}
\dot{E}_{k1 ejected}\,=\frac{E_{k_1}}{t}= 3.17\cdot MW
(\frac{E_t}{10^{17} joule})\cdot(\frac{\eta}{1})\cdot(\frac{t}{10
y})^{-1}
\end{equation}

By geometry, we know that the fraction of radius of the asteroid
skin eroded by mass expulsion, is on average
$\frac{r}{R}=\frac{1}{3}\frac{m}{M}=0.66\cdot
10^{-5}\cdot(\frac{M}{10^{15} g})(\frac{d}{5 \cdot
R_\oplus})^2(\frac{E_t}{10^{22}erg})^{-1}(\frac{\eta}{1})^{-1}(\frac{t}{10
y})^{-2}$

Each one of the, let's say a dozen, screw-engine may eject a smaller
fraction $\simeq 10\%-20\%$ of the power above. The usually huge
amount of expelled  mass (assuming to be delivered to the asteroid
by a chemical rockets) is too large (tens of thousands tons) and too
risky (also for mini-meteorites impact) to be, in our opinion,
seriously considered (nevertheless, see a different opinion in
\cite{Schweickart2003}). Nuclear engines may provide the main energy
at a much small weight (a total few tons range) and the asteroid
mass soil the main propeller thrust (at tens thousand of tons
range).  The wide spread in unities components and moreover their
different places, guarantee the stability and the redundance of the
mission.

The total consequent ejected mass rate is :
\begin{equation}
{\dot{m}}_{ejected}=\frac{2\cdot M^2 d^2}{E_t \cdot\eta\cdot
t^3}=64.54\,g\,s^{-1}(\frac{M}{10^{15}g})^2(\frac{d}{5\cdot
R_\oplus})^2(\frac{E_t}{10^{22}erg})^{-1}(\frac{\eta}{1})^{-1}(\frac{t}{10
y})^{-3}
\end{equation}

Its should be noted the \emph{inverse cubic dependence} with time:
the faster is the needed deflection, the higher is the mass rate
ejection (as well as, \emph{inverse quadratically} total mass
needed.

One may wish to reverse the above formula to derive a needed time
for a deflection of a $km^3$ asteroid. Its characteristic value is,
obviously, already calibrated to a decade. For the same power and
energy in case of ``light'' Apophis the result becomes:
\begin{equation}
t=0.737yr\sqrt[3]{(\frac{\dot{m}_{eject.}}{64.5
gs^{-1}})^{-1}(\frac{M_{Apoph}}{2\cdot10^{13}g})^2(\frac{d}{5R_\oplus})^2(\frac{E_T}{10^{22}erg})^{-1}(\frac{\eta}{1})^{-1}}
\end{equation}

The needed time to reach the object in its periodic trajectory and
to land the array and to make it active should be around few years.

\section{Mass propulsion Vs Radiation pressure.}
To better appreciate the efficiency of the mass ejection over
\textit{massless} explosive radiation (at same energy budget, see
also Appendix A,B), it is worth-full to remind that the momentum
exchange $\Delta P$ for a mass $m$ ejected at velocity $v_1$ is
$$\Delta P=\sqrt{2 E_t m \eta}$$ where $\eta=\frac{E_{k1}}{E_{t}}$ is the total energy conversion into propeller
kinetic one. Note that the above expression may be derived from both
non relativistic and ultra-relativistic regime. Therefore the
adimensional ratio R between the momentum exchange for the present
screw model, with the help of a mass $m$  ejected over an analogous
due to radiation pressure (external nuclear explosions) with albedo
$\hat{\eta}$ at same energy budget, is:

\begin{equation}
R=\sqrt{\frac{2mc^2\eta}{E\hat{\eta}}}=6.32\cdot
10^4\sqrt{(\frac{m}{2\cdot
10^{10}g})}\sqrt{\frac{10^{22}erg}{E_t}}\cdot\sqrt{\frac{\eta}{\hat{\eta}}}
\end{equation}

Therefore, contrary to much written in popular science, to eject
matter is much more effective (nearly five order of  magnitude, as
in example above) in deflecting a $km^3$ asteroid than any radiating
propeller \cite{Fargion 1998}. External radiating atomic bomb are
uneffective (see Appendix A and B). The prompt explosion of nuclear
bombs on the asteroid \emph{interior} may be as good as the screw
array, but this prompt event, even if it is well projected, may lead
to uncontrolled breaking into undesired asteroid fragments
\cite{Fargion 1998}.

Moreover we would reconsider the screw-engine landing and mining the
asteroid. Because of this low gravity the vertical propeller engine
has two role: first, is to eject mass and to thrust the asteroid,
second is to force the engine-screw tank toward the asteroid
surface. Indeed while digging or mining the surface there would be a
very problematic impulse reaction back from the asteroid to the
screw engine. For instance, an astronaut hitting with a fist of few
joules on the asteroid ground could be sent in orbit or at infinity.
Consequently we like to have a persistent engine pressure on the
ground that guarantes a permanent adherence on surface, even while
the screw is digging the soil (within limited pressure). Once again
this force might be spread on each active engine all over the
surface, leading to a tiny $\simeq64 N$ force, corresponding about
to a terrestrial  $6.4$ kg weight and (over a  $\simeq m^2$ engine
base area) a pressure as small as $6.4\cdot10^{2}\,Pa$ (0.6\% of
terrestrial atmospheric one. This thrust is well above the YORP
(Yarkovsky-O'Keefe-Radzievskii-Paddack) effect phenomenon and a
solar light net force (at 1 AU) on a $km^2$ asteroid surface, which
exert a force (even for maximal albedo) higher of a few Newton.
Therefore this Jet activity may rule all the predictable
disturbances by nearly three order of magnitude. It should be noted
that, if the asteroid is made off of pile or pieces as in a rubble
pile model, than such an unit force may accelerate any smallest
finite zone (let us imagine $\sim0.001\%$ of the whole $km^3$
asteroid, an island block of  $10^{10}g$) transferring by gravity
the thrust pull, as in the gravity tractor proposal \cite{Lu T. et
al 2005}. Such a tiniest $10^{10}g$ fragment (let say a sphere)
attracts indeed the other asteroid masses by gravity with an
acceleration

\begin{equation}
\tilde{g}=G\cdot(\frac{4\cdot\pi\rho}{3})^{\frac{2}{3}}\cdot
m^\frac{1}{3}\simeq7.06\cdot10^{-4}cm
s^{-2}\cdot(\frac{m}{10^{10}g})^\frac{1}{3}\cdot(\frac{\rho}{2.6})^\frac{2}{3}
\end{equation}
thousands of times larger that the one produced on the whole body by
the desired deflection (see equation\ref{1}).  The body will move as
an unique object. The eventual but extremely improbable deformation
of a fragile asteroid surface under such a tiny pressure, can be
just overcome by driving the screw-tractor elsewhere or by a
physical linking the screw-engine array into a self-dragging  spread
web-net or wide area wheel extension. Indeed, elastic wheels should
be inflated into a tubular cylinders array (long and extended),
possibly linked in a tractor (like an armored tank) structure. The
wheel area will spread the force and lower the pressure
successfully. Anyway the free-fall or collapse on such a small
gravity body is so slow (hours long) that the engines will have time
to move safely. A preventive soil structure study may avoid such a
surprise by the help of mini-test impact hit.

\subsection{Binary asteroid companion: will it be lost?}

The large presence of multiple or binary companion along asteroids
may lead to a question abou a safe acceleration in such binary case.
As in the previous case, the slow acceleration considered for safe
deflection ${a_{Asteroid}}=\frac{2d}{t^2}=6.4\cdot10^{-8}cm
s^{-2}(\frac{d}{5\cdot R_{\oplus}})(\frac{t}{10 y})^{-2}$, is well
below the characteristic gravity on the surface; for sake of
simplicity let us imagine a spherical $km^3$ body): ${g_{Asteroid}}=
G(\frac{4\pi\rho}{3})^{\frac{2}{3}}\cdot
M^{\frac{1}{3}}=3.28\cdot10^{-2}cm s^{-2}\cdot(\frac{M}{10^{15}
g})^{\frac{1}{3}}$. It comes out that the dragging of the heaviest
primary asteroid will smoothly and safely drive also the second
companion as a unique object.

\section{The asteroid spin role.}

A $km^3$ asteroid is in general an irregular shaped body. Because of
it the sun light pressure acts a net torque on the asteroid. Because
of such a YORP phenomena (Yarkovsky-O'Keefe-Radzievskii-Paddack
effect) it has been early estimated  (and also recently observed on
1862 Apollo spin rate \cite{Mikko 2007}) that the asteroid rotation
may be propelled and accelerated to the maximal angular velocity
(just before to break apart). In general the angular velocity is non
zero. Let us consider as first estimate a spherical prototype whose
own gravity and inertial momentum is well predictable: we may
estimate its own kinetic rotational energy (at critical break
angular velocity or even just at a smaller Keplerian  one) and
compare it with the desired asteroid center of mass kinetic
displacement. It is easy to show that for such critical (orbital)
spinning asteroid, its rotational energy $E_{Ast-Rot}$ is:
\begin{equation}
{E}_{Ast-Rot}=\frac{1}{5}\cdot
M^{\frac{5}{3}}\cdot(\frac{4\pi\rho}{3})^\frac{1}{3}\cdot
G=2.96\cdot10^{17}erg\cdot(\frac{M}{10^{15}
g})^\frac{5}{3}\cdot(\frac{\rho}{2.6})^\frac{1}{3}
\end{equation}

This value is comparable to the asteroid kinetic deflection energy
shown before
$$E_{k Ast}=2.04\cdot10^{17}erg\cdot(\frac{d}{5\cdot R_{\oplus}})^2(\frac{t}{10y})^{-2}(\frac{M}{{10^{15}g}})$$
and it implies an additional (at least a $\frac{5}{2}$ larger
factor) energy cost and a more difficult control as in any one
unique tug (\cite{Schweickart et al. 2003}) engine project. Moreover
because  ${E}_{Ast-Rot}\propto M^{\frac{5}{3}}$ while
${E}_{Ast-c.m.}\propto M$ it is obvious that for large mass
asteroids $\geq1.5-2 km^3$ the de-spin energy request will largely
exceed the deflection one. In a different and popular scenario where
the deflection is due to the pulling by a sail (many square $km^2$
area) one faces the unavoidable problem for almost all spinning
bodies, on how and where to held the thrust on such turning massive
body. If we don't have stability along the principal axis it will
turn out to be an almost impossible mission: there is not in general
any fixed tracking point where to hang the cable to the sail (or
external jet engine). We suggest to overcome the asteroid rotation
disturbance by a cooperative engine action on the asteroid surface,
mainly in three different (but  complementary) procedures that we
summarize here as: \emph{A tuned Phase Array activity,
Contra-Rotating Procession Array and a turning Sunflower or
Earth-flower  array}.

Let us first briefly describe the unity engine, the preliminary
test, its parental missile structure and the individual engine
landing procedure on the asteroid.

\subsection{Deflecting a spinning Asteroid: Preliminary test.}

Any landing on asteroids of the main rocket array, must be preceded
by precursor mini-stationary satellites whose role is to be on orbit
and to inspect the soil nature, composition and structure. Assuming
a few AU distances of the asteroid position, the whole preparation
(first test landing, final array landing, deflection) for a $km^3$
size, might take a decade.  Different tracking exploration will
guarantee the success. The first landing of mini acoustical, radio,
gamma and optical devices might better characterize the asteroid
inner cohesion, as well as its exact morphological map, gravity
potential and spinning behavior. An array of light retro-reflectors,
mirrors and antennas may offer at best echoes tracking.
Piezoelectric acoustic emitters and detectors may offer a 3D
physical test of the asteroid. This preliminary study will leave  an
independent radio array to better communicate and a  gyroscope array
on the surface and on nearby asteroid-stationary  orbit to track the
spinning details of the body. This may be the case of Apophis,
discussed also in section \ref{a} and in appendix \ref{F}.

\subsection{In flight from the Earth to an incoming Asteroid.}

The flight and the landing might take place, if possible, either in
one of the eventual nearby asteroid periodic Earth encounter, or
viceversa for any one-way incoming  impact event, by a fast reaching
to the asteroid in flight, possibly accelerated  by other planet's
gravity bending and swinging as well as by a rapid slowing down into
stationary orbit and on the soil, at lowest Kepler orbital
velocities (a fraction of $m s^{-1}$); for a spherical ideal case
(that maybe simply estimated) this Kepler orbital velocity is:
\begin{equation}
{v}_{Ast-Kepl}=G^{\frac{1}{2}}\cdot M^{\frac{1}{3}}\cdot
{\rho}^{\frac{1}{6}}\cdot({\frac{3}{4\pi}})^{(-\frac{1}{6})}=38.45
cm s^{-1}\cdot ({\frac{M}{10^{15}g}})^{\frac{1}{3}} \cdot
({\frac{\rho}{2.6}})^\frac{1}{6}
\end{equation}

\subsection{The Landing of Unit Screw-Rocket engine in a safe oval Airbag.}
We suggest to build a robotic unit (of the whole array) made as a
mini ``screw'' tank of a few hundred $kg$ mass or less, based on a
screw-rocket structure, whose mini-nuclear engine is digging and
drilling downward the soil while being ejecting it at high speed
(ten $km s^{-1}$),  upward into a (nearly vertical) jet beam. Such a
mini nuclear engine should not be confused  with much larger and
heavier nuclear thermal rocket considered elsewhere for
self-sustained lunch or longest planetary voyages. The center of
mass of the oval airbag, guarantees a soft rolling and a vertical
standing (because of low gravity
${g_{Asteroid}}=(G\frac{4\pi\rho}{3})^{\frac{2}{3}}\cdot
M^{\frac{1}{3}}=3.277\cdot10^{-2}cm
s^{-2}\cdot(\frac{M}{10^{15}g})^{\frac{1}{3}}\cdot\frac{\rho}{2.6}^{\frac{2}{3}}$)
in a very slow oscillatory period (assuming a three meter oval
height):

$$P= 601.1 s \cdot(\frac{M}{10^{15} g})^{(-\frac{1}{6})}(\frac{\rho}{2.6})^{(-\frac{1}{3})}(\frac{h}{3 m})^{\frac{1}{2}}$$

Therefore the relaxation time is long and the vertical engine
propulsion may increase the apparent gravity and the speed up
relaxation time. The common chemical boosters might be  the main
carrier rocket (containing the payload of a dozen of screw-rocket
unities, possibly located in two ore three independent landing
rocket heads). The whole parental missile  will depart from the
Earth surface by the usual chemical thrust engines, while in outer
spaces it will be partially accelerated to high speed and
decelerated by additional (few tons)  power-full  engines (this one
using on board chemical propeller and eventual nuclear thruster).

\begin{figure}[t]
\centering
\hspace{30pt}\psfig{file=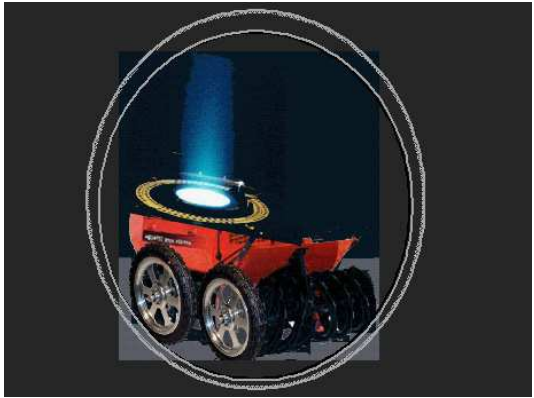,width=7cm}
\caption{The multi-layer oval envelope that will be inflated around
the screw-engine at the landing time, offering a safe bouncing,
rolling and a final standing up on the asteroid soil. The egg-like
bag, possibly made by twin-vertical slice component, will be safely
abandoned on the ground or deflated on the engine sides.}
\label{fig1}
\end{figure}

The whole array and the parental courier engines (that will remain
for control and communication in asteroid orbit) may reach a mass of
nearly ten tons. Just to compare the lunar payload for the various
past missions, it varied between 48 and 75 tons. The asteroid
landing will occur at very low spiralling orbital speed, as small as
a fraction of $m s^{-1}$. Because of it there is not need of any
delicate landing procedure but just a few safe air-bags or twin
halves of egg-like inflated airbags: they may allow a smooth rolling
and bouncing and landing along the asteroid, leaving the engine
standing vertically on asteroid surface. The expulsion of such
mini-airbag, or the undressing of the egg-like landing envelope may
take place once at rest at vertical position. The procedure is
simple and testable on Earth (see Fig.\ref{fig1}).

\vspace{4pt}
\begin{figure}[]
\centering\hspace{30pt}
\psfig{file=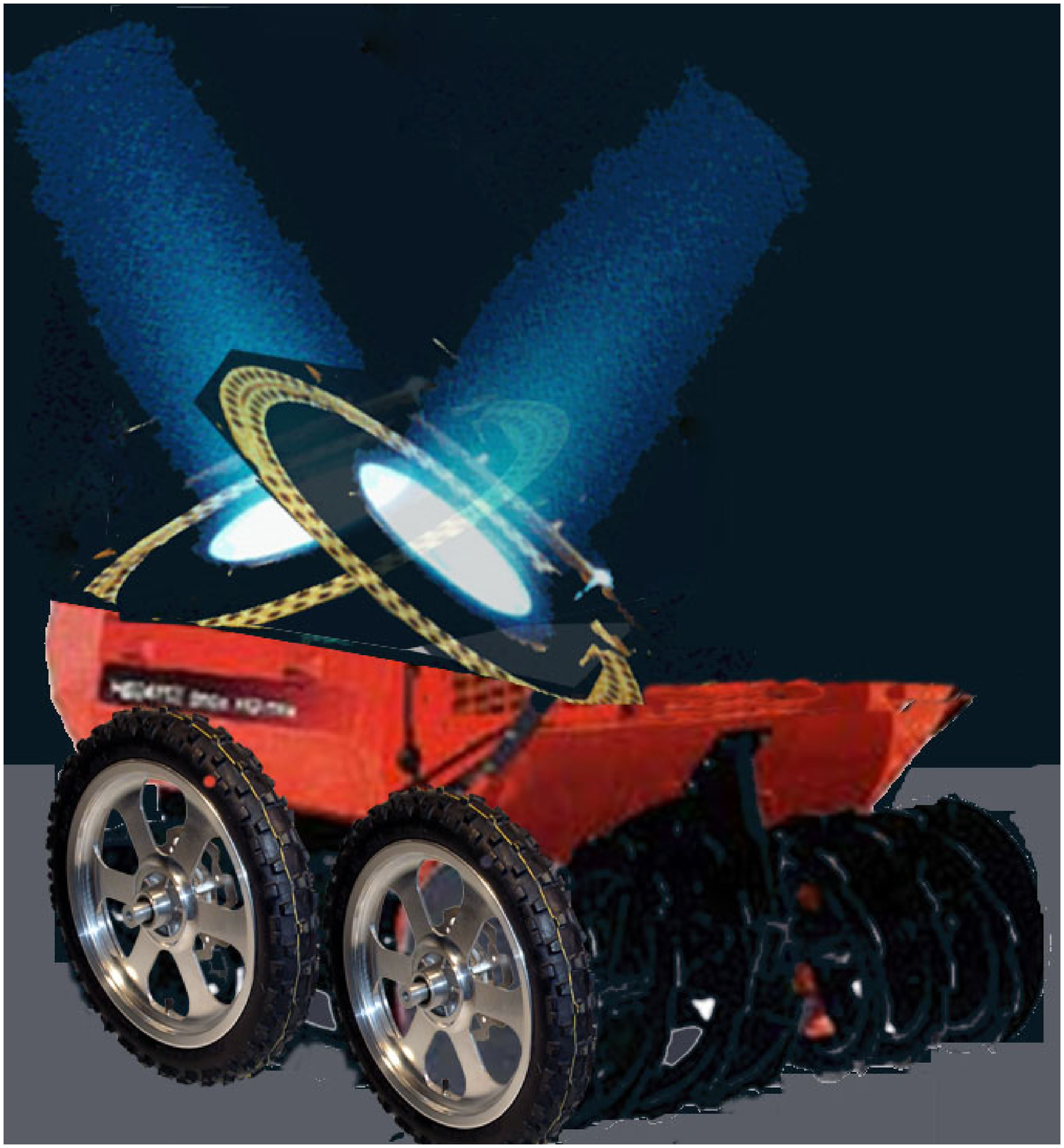,width=6cm}
\caption{The screw-engine tractor might bend partially its jet
directions, leading to a wide angle sky coverture. This angle
(related to the friction coefficient $\mu$) is:
$\theta\leq\arctan\mu$). It may guarantee the inclined emission
within a narrow or wide solid angle $\Omega=\pi\theta^2$).}
\label{fig2}
\end{figure}
The screw and dig procedure occurs while the engine is moving along
its surface. Because of the asteroid extremely low gravity and the
risk of inclined escape from the surface, we imagine a cooperative
ejection (of the rocket) orthogonally to the ground, whose reaction
and thrust guarantee downward pressure and complete adhesion on its
asteroid surface. The landing at extremely low gravity is offered by
a slow (a fraction of $m s^{-1}$) rolling of the engine inside an
inflated spherical multi-layer, possibly a transparent air-bag
envelope, as suggest by figure above.

\subsection{The simplest, unrealistic, non rotating asteroid.}

In a very rare case  the asteroid is not rotating respect to the
Earth; in such an ideal situation, the engines location may face
(respect to us) always the same side. In this  peculiar case there
is a point or a slow moving region on the surface (defined by the
intersection Earth-Asteroid) where to locate the propeller engines.
This eventuality could be faced by an ideal array configuration
clustered in a very narrow area as described, for instance, in
Fig.\ref{fig3}. There is also a very tiny, though non zero,
probability that the asteroid principle inertial momentum axis and
spin is located at the main symmetry axis and mostly collimated
toward (or opposite to) the Earth.  As before, in this very
fortuitous case, the deflection maybe offered also by a few well
located engines. But there are good reasons  not to relay on this
unprobable case.

\subsection{To compensate the asteroid  spin: \textbf{\emph{A tuned Phase Array activity}}.}

A wide and wild spinning of the asteroid, may need a random spread
of screw rocket engines on the asteroid surface; these rockets will
switch their engines at given synchronous times in analogy of our
Earth human cities, that switch on the lights while in the night
time: so the rocket-screw engines will act every time is facing, for
instance, the Earth (as in the simplest case).

\begin{figure}[]
\centering\hspace{30pt}
\psfig{file=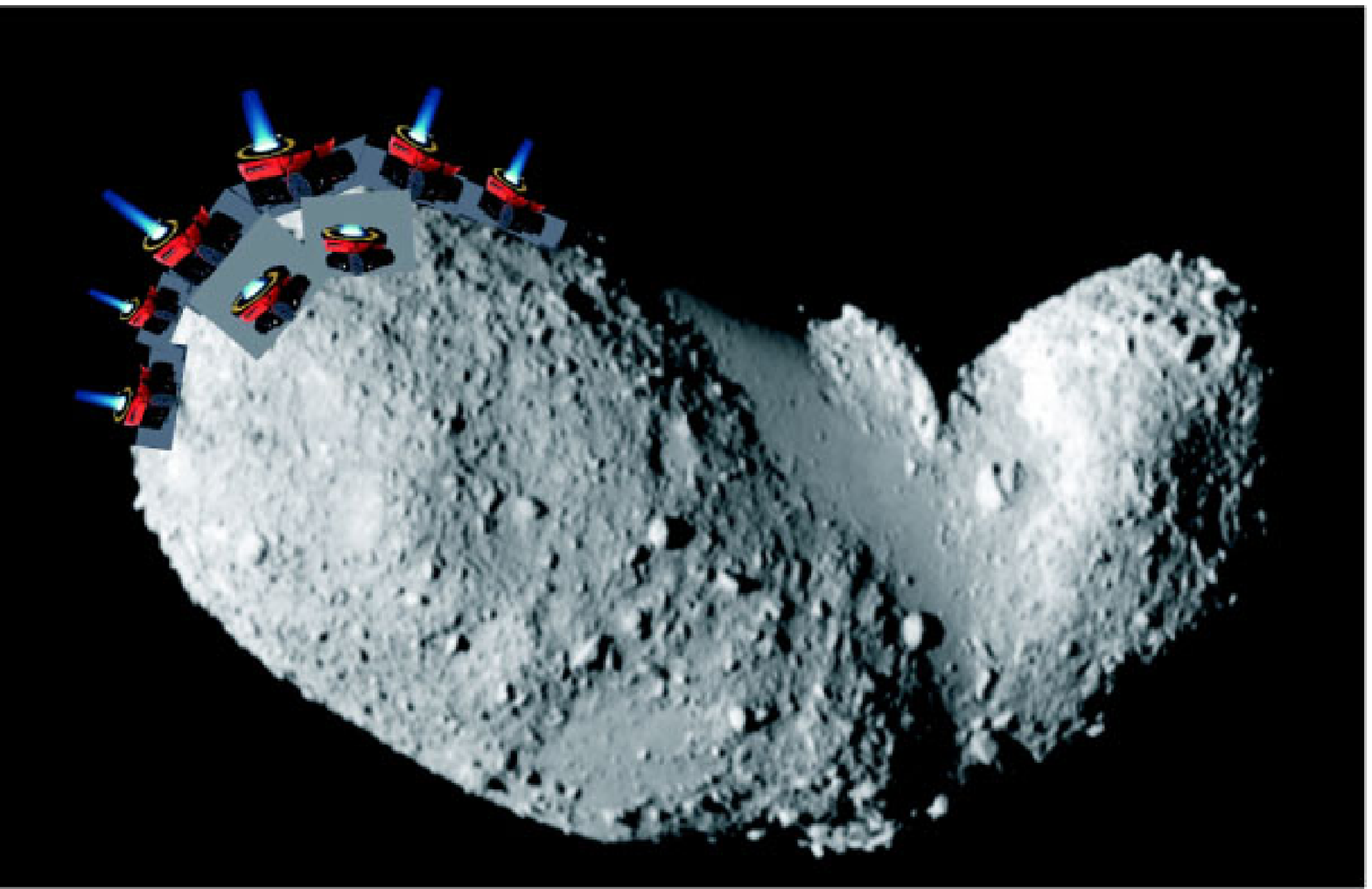,width=7cm}
\caption{A non rotating or a spinning asteroid in axis toward the
Earth, may offer the simplest solution of a clustering of a few
screw-engine in a well defined area, as described in figure. In
principle it may exist just a point on the surface, connecting a
line from the Earth toward the asteroid center of mass intersecting
the body surface; however, even in the most optimistic
configuration, this point by slow orbital spin will define at best a
small  area where to locate the engines. } \label{fig3}
\centering\hspace{30pt}
\psfig{file=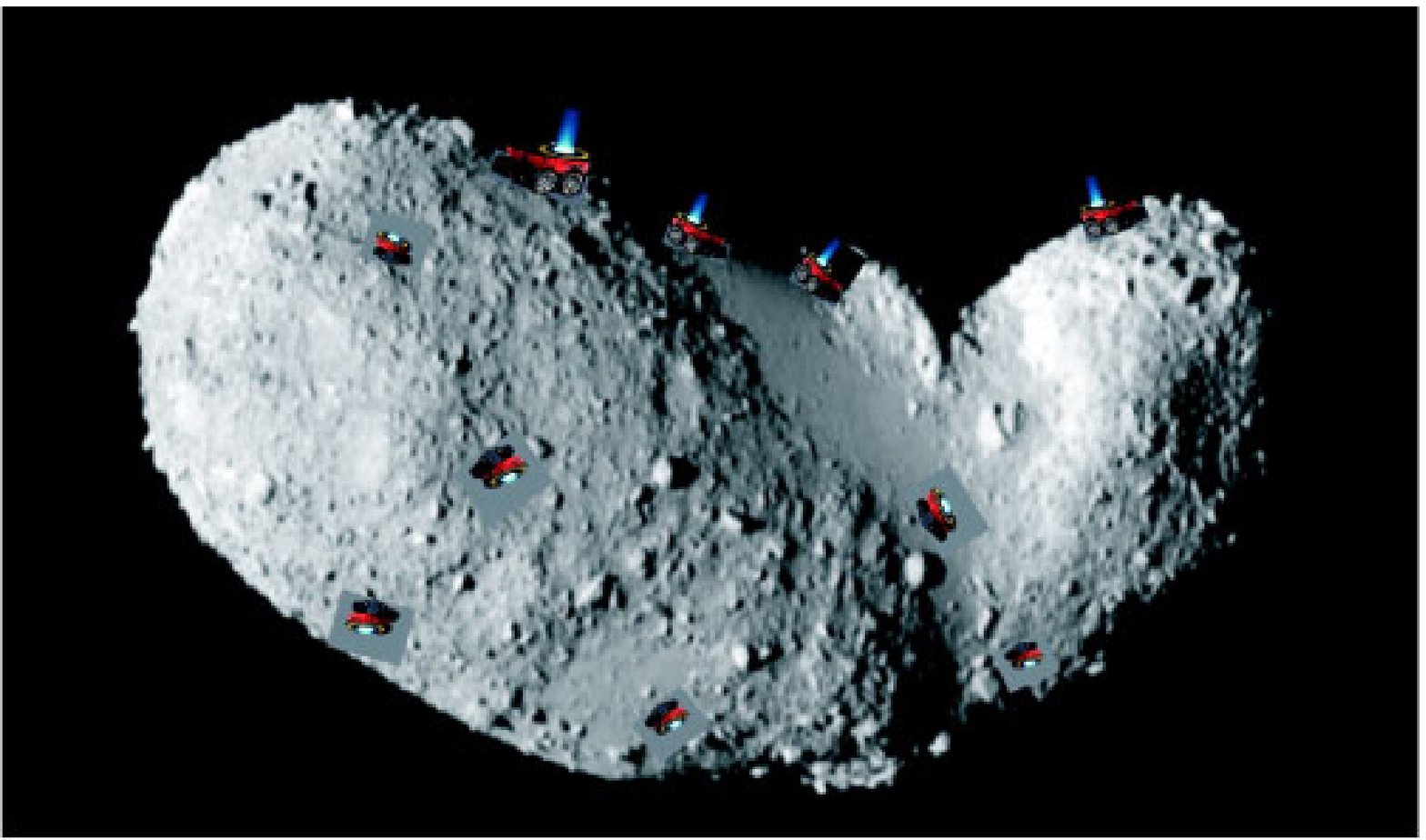,width=7cm}
\caption{A number of screw-engine tractor might activate the jet in
phase compensating its rotation; an additional bending of its
directions is leading to a better coherent thrust. The total number
required is  related to complete sky coverture. The bending angle
(related to the friction coefficient $\mu$) is:
$\theta\leq\arctan\mu$ and it depends on the screw-engine material
over the asteroid one). It may guarantee the inclined emission
within a solid angle ($\Omega=\pi\theta^2$).} \label{fig4}
\centering\hspace{30pt}
\psfig{file=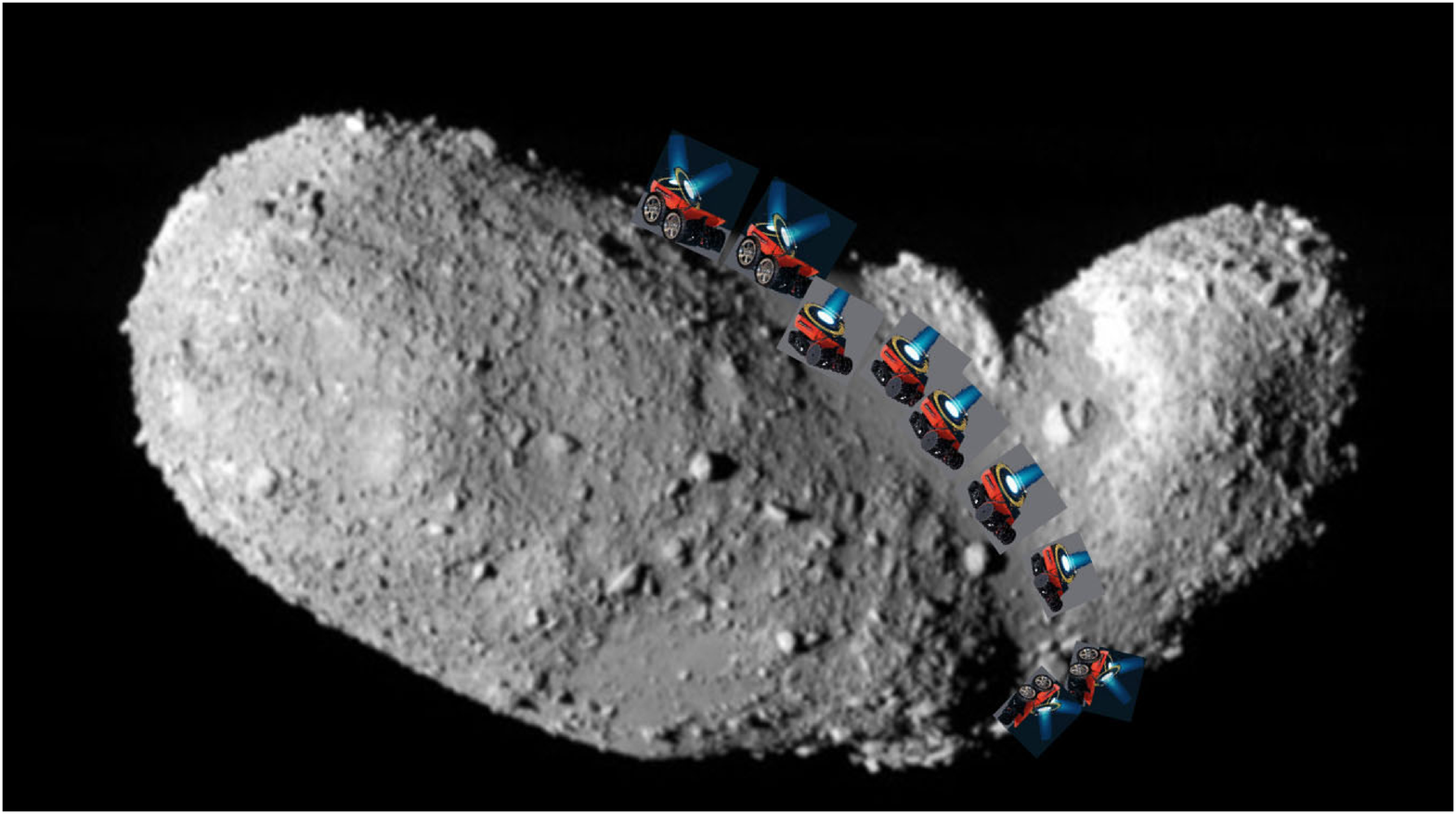,width=7cm}
\caption{A procession of the screw-engine  jet leading to a coherent
thrust contra-rotating and vanishing the asteroid spin.}
\label{fig5}
\end{figure}

The multi jet action at synchronous phase will cope with the
asteroid rotation and it will push coherently to a needed direction.
This procedure imply dead time for most of the engines and it may
reduce the whole thrust efficiency. However, the engines reside
within their nearby landing places and do not  move much far away on
the asteroid surface. Moreover, the estimated energy of thrust
engines may be considered as an average one: therefore the whole
mass and energy output in above equations remain the same ones. The
eventual need of  contemporaneous ejection of a number inclined
engines-jet at different angle $\alpha$ may reduce by a factor
$cos(\alpha)$ the whole output efficiency. For a maximal spread of
$\Delta\alpha=60^\circ$ the efficiency will be within a factor of a
half. Fig.\ref{fig4} describes the possible set up of a large array
distributed either at will or in a random way. The possible turning
of the engine beam jet pointing (as Sun-flower or better as an
\emph{Earth-Flower}) makes the efficiency larger while a partial
engine array procession, discussed below,  might also increase the
efficiency back to the unit (see Fig.\ref{fig2}).

\subsection{A \textbf{\emph{Contra-rotating}} Procession array.}
In the case of a safe mobility on the asteroid, the
mini-screw-engines may follow or trace a \emph{road-map} on the
asteroid surface whose trajectory is defined by the intersection of
the line from the Earth to the Asteroid center of mass, with the
asteroid surface. This \emph{road-map} might be a point (or a small
circle) on the body tip if the asteroid is spinning along a
principal axis who is, at the same time, pointing toward the Earth,
as discussed above. In general it may be a ring or arrays of rings
defining a \emph{road-map} possibly never returning at the same
starting points. A slow procession of the screw engines may always
stand in the optimal place in axis with Earth-C.M. (Center of Mass
of the Asteroid, see Fig\ref{fig€5}).

\subsection{\textbf{\emph{A turning Sunflower or Earth-flower jet array}}.}

If the surface of the asteroid with the screw engines, and its
tires,  offer a large friction coefficient (as on terrestrial soil),
$\mu \simeq 0.5-1 $ , than the jet propulsion may be turned not just
vertically to the surface, but also at moderate inclined zenith
angle: $\theta \leq arctg \mu $. For the values $\mu = 0.5-1 $ the
consequent sky coverture by jet bending becomes:
$\frac{\Delta\Omega}{\Omega} = \frac{\pi (\arctan \mu)^2}{4 \pi}
\simeq    5.4 \%$ and $ 15.5\%$ of all the sky or at least $\Omega
\geq 0.679 sr.$ The inverse of these fractions (respectively $18.5$
and $6.45$ defines the number of unities to complete $4 \pi$ sky
coverture.  In general a dozen of engines may be a quite reasonable
number. It is the jet bending (for instance as an Earth flower mode)
that may thrust the asteroid leading to the persistency of the array
unity all over the flight; an additional slow contra-rotating
procession (as prescribed above) may better optimize the cooperative
action of the array (see Fig\ref{fig2}).

\subsection{The first useful mining on the Moon}
The test of robotic screw-tractors on  the Moon is a first goal:
such a robotic array on lunar soil may mine and  provide first
galleries and tunnels for human station and shielding  for permanent
biophysical and human scientific laboratories. The eventual small
nuclear pollution on empty moon will not induce relevant problems;
the quality test for the automated system will be a very important
step for  a permanent landing and stay on our natural satellite.

\subsection{A pro-po of 99942 Apophis and Swing Gravity: A Hazard or an
Opportunity?}\label{a}

(99942) Apophis (previously known by its provisional designation
2004 MN4) is a near-Earth asteroid; in future is expected to reach
an impact Earth distance on $2029$ of $0.0002318$ Astronomic Unit
($34.700 km$) forcing to a question of hazard and urgency, calling
for the feasibility of  needed deflection. The above formulas have
been advocated for a $km^3$ rarest masses, while the Apophis one is
much smaller, nearly $2 \cdot 10^{13} g$, or $50$ times less the
considered unite mass. Therefore the deflection ability considered
for $km^3$ case is nearly $2500$ the needed one for Apophis and the
mass request may be reduced to a smaller fraction (just a few  tens
tons). This mass maybe evacuated by micro nuclear
screw-engines,whose weight may be extremely light. A few terrestrial
radius deflection means a great opportunity: to use the Earth (or
Venus or any planet) gravity swing at our will.  Indeed the gravity
swing (even its General Relativity extension) may amplify the future
trajectory by wide spreading angle \cite{Fargion 1981}:
$$\Delta\varphi =  2\cdot \frac{G M_\oplus}{c^2} \frac{1+ \beta^2}{b {\beta}^2}$$  where $\beta$ is the incoming-out-coming  asteroid velocity at infinity in $c$ unite, $b$  is the impact distance, $M$ is the bending body. For the Apophis visit this angle value is:
$$\Delta\varphi \simeq  15.2^o \cdot (\frac{v}{10 km s^{-1}})^{-2} (\frac{ 5 R_\otimes}{b}) $$

Because of the large deflection $b$ may be changed by  a factor 2-3
leading to a wide solid angle where the asteroid might be driven.
Assuming a general maximal deflection (at 2.5 Terrestrial radius)
the whole solid angle is $ 7\%$ of the whole sky. A large windows of
opportunity into space.

\section{Conclusions: Amplified asteroids deflection by gravity swing to drive them in space}
A safe and successful asteroid deflection (a few Earth radius) finds
a solution by the landing of an array of screw-robotic engines (at
best of nuclear nature) able to co-work in phase, keeping care of
the body rotation. Such an array maybe used to create a first
gallery net inside the Moon to guest permanent human laboratories.
Exploiting  mini asteroids, driving them by array engines at highest
speed (also aided by planetary gravitational swing) may offer a
novel safe shielding  for future human interplanetary flight. Indeed
the proliferous presence  of incoming smaller asteroids NEO (at 20
meter size) may allow fastest deflections. Indeed for the lighter
the  body (not billions, like $km^3$ but few tens thousands tons)
the larger  is the final velocity achieved. $v_{Asteroid-final} = 10
km/s \frac{v_1}{10 km /s}\ln[\frac{M_i}{M_f}] $; if the final mass
is , for an example, half of the initial one, than the final
velocity maybe $v_{Asteroid-final} = 6.93 km/s \frac{v_1}{10 km /s}
$; therefore it could be possible to drive  the asteroid to a planet
gravity swing able to deflect and accelerate the asteroid to highest
velocities in solar system. It maybe possible to project a \emph{
hijack} of nearby small asteroid  deflectable toward Mars. The inner
(few meters under-ground) spaces of the asteroids may offer a safe
container for biological eco-system and even human room able to
survive longest trip screened by most  dangerous cosmic rays
radiations. Indeed the largest solar flare threat (often taking
place in a few years duration) may lead to a lethal radiation dose
to astronauts. The procedure maybe experienced on nearby Earth
Object (NEO) or on far Jove (a few $Km^3$)   captured asteroids:
S/2003 J12 19002480; S/2003 J9  22441680.

The Apophis near encounter at few terrestrial radius deflection
means a great opportunity: to use the Earth  gravity swing at our
will. It is not yet clear if the deflection may point to Mars, but
probably it maybe forced to Venus. Indeed the gravity swing  may
amplify the future trajectory by wide spreading angle and the
asteroid mass may be the novel niche where life may travel into
inter-planetary spaces. Or at least asteroid mass is a reserve where
to pick up propeller (dust) to feed and fuel spacecraft jet engines.
In this view and because of the low gravity and a low cost docking
on asteroids, it is preferable  to first land on a large asteroids
(tens of thousand $km^3$) like Phobos or on Deimos, than on Mars
itself. There are very good reasons why these  asteroids, may guest
in the inner core the first human station in martian space, from
where to visit, time to time,  the main planet. Therefore small and
large asteroids may be the courier, the shielding, the refueling
station  and the host for new human steps  into deep space.
\subsection{Aknowledgments}
The author wish to thank Dr.Aiello that suggested the subject in
Appendix B, as well as Drs.O. Lanciano, P.Oliva, F.La Monaca  for
useful discussions and  M.Farina, for a  mathematical control.  The
paper is devoted to the memory of  Margherita Habbib Fargion , born
on 12 May 1923, and Yakov Evron Abudi , lost on 1 May 2007.

\section{Appendix A: How many Nukes for an asteroid deflection?}
One would like to compare the needed Screw-engine output
\begin{equation}
\dot{E}_{k1 ejected} \,= \frac{E_{k_1}}{t}= 3.17 \cdot MW
(\frac{E_t}{10^{17} joule}) \cdot (\frac{\eta}{1})\cdot (
\frac{t}{10 y})^{-1}
\end{equation}
as well as its total energy $E_t = 10^{22}erg/ \eta$ (ten Hiroshima
bomb energy or a week of large nuclear, GW power activity) versus a
competitive \emph{external} nuclear blast prompt deflection.

One should  note that radiation being massless, for a same energy
content, offer much less momentum than a mass with same kinetic
energy. Moreover the spherical atomic blast momentum, to be captured
at most, might explode nearby the asteroid: its huge consequent
thermal inhomogeneity and  corresponding momentum impulse
$E_{Hiroshima Bomb}/c = 3.33 \cdot 10^{10} g cm s^{-1}$ might break
down  the body integrity. Anyway the  total needed momentum $\Delta
P_{ Ast} = 2.018 \cdot 10^{16} g \cdot cm s^{-1} (\frac{d}{5\cdot
R_{\oplus}})(\frac{t}{10y})^{-1} \cdot(\frac{M}{10^{15} g})$, might
be reached by $$ N_{Hiroshima Bomb}= 6 \cdot 10^5.$$ Such a huge
explosion might be occurring coherently or might be beamed as a
plane wave, possibly each one, by an ad hoc paraboloid $ \simeq
km^2$ sail-mirror, whose structure, nevertheless will be immediately
evaporated by the explosion. Otherwise  nearly a  million coherent
bombs in half a sphere should explode in phase and time on the
desired asteroid side. This configurations seem unrealistic and
nevertheless very risky because of very probable fragmentation.

\section{Appendix B: Why Neutron Bombs transfer large impulse?}
Neutron bombs, also called enhanced radiation bombs (ER weapons),
are small nuclear or thermonuclear weapons in which the burst of
neutrons generated by the fusion reaction is intentionally not
absorbed inside the weapon, but allowed to escape. Because a large
fraction of the explosion is focused into baryons of mass m (and not
in radiation) the courier momentum is higher than in radiating nukes
(the effect grows with the square root of the mass). The exchanged
momentum is: $\Delta P = \sqrt[2]{2 m E}$ For instance for a neutron
bomb mass m as large as $20 kg$, at  a smaller energy (nearly two
k-ton TNT)  will eject almost $ 2 \cdot 10^{12}$ g cm $s^{-1}$
momentum, nearly a hundred times larger than the previous case.
Moreover the same result maybe enhanced by an order of magnitude if
the ER loaded mass is as large as 2 ton. Such a deflection may  lead
to a prompt deflection speed for incoming Apophis asteroid ($2\%$ of
$km^3$ mass) as large as $1 cm s^{-1}$, corresponding  to a
deflection in a decade scale of nearly $3000 km$ distance. Therefore
even if quite dangerous because of possible asteroid fracture, the
mass loaded nuclear bombs (ER) are an interesting options for fast
deflections. However the neutron-bomb  lifetime is limited because
they are composed by tritium, which has a half-life of 12.3 years.
In such a long journey it maybe a problem or an handicap. However to
make the effect larger one may imagine to use larger asteroid skin
mass,  as indeed in the case of the explosion inside the asteroids
\cite{Fargion 1998}. However with more danger on the asteroid
structure survival. In the same view our present screw-array nuclear
engines are doing the same procedure in a longer but much more
controlled way : for a nominal asteroid  mass ejected $ m =  2 \cdot
10^{10} $ g,  and a total energy $E_t = 10^{22}$ erg, comparable to
ten Hiroshima bombs, for a $km^3$ mass asteroid prototype, we
recover exactly the same result as for screw array considered in the
text.

\section{Appendix C: May Asteroid be charged and deflected  by  Lorenz Forces ?}
It is in principle interesting to consider the deflection of an
asteroid by charging its surface (by ionic emission) and by bending
its trajectory because coherent solar magnetic field.  This
procedure , while being elegant do not guarantee the needed bending
as well as the solar magnetic field coherence. Nevertheless just to
have a first estimate we compare the needed Charge for a force with
the maximal Lorenz one derived in maximal solar field:
$$
Q_{Asteroid} \cdot B_{Solar}\cdot \overline{v}_{Asteroid}=
{F}_{Asteroid}= M \cdot a_{Asteroid} $$
$$=2 \cdot M \cdot d  \cdot t^{-2}= 640.5 N (\frac{M}{10^{15}
g})\cdot (\frac{d}{5 \cdot R_\oplus})\cdot (\frac{t}{10y})^{-2}
$$
From here one derives:
\begin{equation}
Q_{Asteroid}= 6.4 \cdot 10^6 C (\frac{M}{10^{15}g}) \cdot
(\frac{d}{5 \cdot R_\oplus})\cdot
(\frac{t}{10y})^{-2}(\frac{B_\odot}{10^{-4}
Gauss})^{-1}(\frac{\overline{v}_{Asteroid}}{10 km s^{-1}})^{-1}
\end{equation}
The charge will lead to a highest potential on the Asteroid surface
(here assumed spherical) whose capacity is $ C = 4 \pi \epsilon
R_{Asteroid} \simeq 5.17\cdot 10^{-8} \cdot
\sqrt[3]{\frac{M}{10^{15} g}}({\frac{\rho}{2.6 \cdot 10^3 kg \cdot
m^{-3}})^{-\frac{1}{3}}}$;

$$\Delta V = (\Delta Q)/C \simeq 1.23 \cdot 10^{14} (\frac{M}{10^{15}g})^{\frac{2}{3}} \cdot (\frac{d}{5 \cdot R_\oplus})\cdot (\frac{t}{10y})^{-2}(\frac{B_\odot}{10^{-4} Gauss})^{-1}(\frac{\overline{v}_{Asteroid}}{10 km s^{-1}})^{-1}\cdot  V $$.
The needed energy to charge such a huge net charge over the asteroid
is prohibitive:
\begin{equation}
E_{Charge-Asteroid}= 3.9 \cdot 10^{27} erg
(\frac{M}{10^{15}g})^{\frac{5}{3}}\cdot(\frac{d}{5\cdot R_\oplus})^2
\cdot (\frac{t}{10y})^{-4}(\frac{B_\odot}{10^{-4}
Gauss})^{-2}(\frac{\overline{v}_{Asteroid}}{10 km s^{-1}})^{-2}
\end{equation}
This extreme values is nearly $4 \cdot 10^ 6$ Hiroshima bomb energy
and it does'nt guarantee the needed deflection road; indeed one may
imagine, in principle,  a controlled charge change along the
asteroid trip, as well as a verified and controlled solar magnetic
field map, at each stages. It is  surprising and exciting to imagine
the electromagnetism dominance over the gravitational trajectory of
a massive body. Because of the energy $E_{Charge-Asteroid}$
dependence on the asteroid mass exponent, $\frac{5}{3}$, versus the
linear dependence in screw-array procedure, at lowest masses (a few
hundred tons asteroids) the procedure maybe in principle of
interest. But such bodies are already safely evaporated in
atmosphere. Indeed at lightest edges, charged particles like  cosmic
rays follows the Lorentz forces more than gravity ones.

\section{Appendix D: Are Anti-Asteroids anomalous deflections  observable?}
In our Universe and also in our galaxy and solar system there might
be present a tiny relic antimatter trace.  There are ongoing
experiment (AMS) in space looking for primordial anti-nucleolus
relic in cosmic rays \cite{AMS Collaboration 2007}.  The possibility
that such larger block, as anti-meteorites, maybe crossing our space
and hit Earth, Moon , Sun and  planets has been widely investigated
\cite{Fargion-Khlopov 2003}. This possibility is strongly
constrained by the  absence of $\gamma$ mini-burst on Moon, Earth,
Jove and Sun. However  there are on-going  projects to detect
eventual anti-helium in cosmic rays. In this view  there is an
additional argument regarding the eventual rare  presence of
anti-asteroid   (mostly mini-ones) whose trajectory in solar system
maybe disturbed by gas matter annihilation on antimatter  along
their flight trajectory. Even if the amount of energy annihilated at
$km^2$ in ten year period is comparable to the one considered to
power the screw-engines   the final momentum exchange is negligible:
this is because the radiation pressure is much   less effective in
making force than mass expulsion. Indeed the annihilating nucleons
are   producing four-five ultra-relativistic pions whose behavior is
comparable to massless radiation. Being the effect of annihilation
proportional with the surface the total effect is independent on
size; however along  solar or planet's  atmosphere the higher gas
density and the anti-matter higher annihilation leads to observable
enhanced repulsion or bouncing  ruling the antimatter trajectory
\cite{Fargion-Khlopov 2003}. Therefore one cannot use eventual
anomalous asteroid trajectory in solar system to infer or speculate
on their eventual anti-matter nature.

\section{Appendix E: May fast Spinning  Asteroid  broken down and split in pairs?}

The existence of fast spinning asteroids, whose rotational energy
may exceed the total kinetic one  for a complete deflection, inspire
a process that convert this ready spinning energy into a
translational kinetic one. The process, eventually aided by a
central breaking explosion, may divert two main fragment
approximately at their   extreme rotational speed (tens of cm/s),
converting part of the present rotational energy  into prompt twin
kinetic ones  (just  a few tens of cm/s). In the most urgent
eventuality  it is  possible to induce  such  a  splitting
deflection, by the same screw-engine array  thrust and torques,
spinning up the body to its maximal angular velocity. In an ideal
longitudinal asteroid the break event might be also induced by a
central nuclear  belt array  coherently exploding. Such a twin
fragment deflection while being the fastest solution it is still a
very risky process: the eventual multi fragment and the uncontrolled
fragment size break and velocity may generate undesired
consequences. It is nevertheless the fastest deflection that in some
extreme case might be taken into consideration.    An interesting
speed up maybe obtained in slower, but cheaper and coherent way by
an asymmetric painting of the body, exploiting the solar radiation
pressure ( see ref.\cite{Lowry07}) and the YORP phenomena
(Yarkovsky-O'Keefe-Radzievskii-Paddack effect).

\section{Appendix F: How to tag  Apophis  trajectory on 2017}\label{F}
One of the simplest and cheapest way to follow the Apophis
trajectory with high accuracy is to land or to put in its orbit a
precise clock-radio-transmitter: its battery will be fed by solar
array, its output will economize by bursting  emission (via beamed
array) to Earth direction in synchronous coherent radio-wave
(nominal duration 100 nanosecond); this timing , like in Pollicino
tale story, in a precise correlation with a terrestrial twin clock,
will mark the asteroid distance ($z$ depth) from Earth. Every lapse
period (hours) will trace the  depth distance with necessary
accuracy:  indeed the present cesium (or even better) atomic clock
may tag the time   within an error below a second every million
years. Such a timing accuracy may guarantee  a distance of  an error
dispersion (for a decade time) of 10 microseconds, corresponding to
a distance below $3 km$, offering better than three sigma
longitudinal distance evaluation ($12 Km$). However the relativistic
delay dilution due to radio-clock traveling velocity $\simeq
\frac{\beta_{Asteroid-\oplus}^2}{2} \simeq 8\cdot 10^{-8} $ must be
estimated, counted out, or emulated  and subtracted; for instance by
twin clock in peculiar comparable trajectories  (for instance in
tuned Jove orbit). To obtain a precise angular resolution the
signal, for instance at 30 Gigahertz band,  might be recorded, by
interference, from largest allowable edges. Simultaneous observation
of a radio source with the Halca satellite and the ground VLBI
network makes  possible to obtain images with the same angular
resolution as images obtained with a single dish as large as the
maximum distance. This distance is about twice the Earth diameter.
The consequent angular resolution $\leq  10^{-9} $ rad, makes easy
to tracks within $\ll km$ the Apophis transverse distance ($x-y$),
offering as for the depth $z$ an accurate track of the asteroid
position.

A better depth tag $z$ maybe achieved via echoes timing, where the
clock radio tag emit an automatic burst message  replying to a
trigger terrestrial one. Half of the two way time multiplied the
velocity of light define the Asteroid distance. No relativistic
correction are needed. In this scenario however the radio-clock
should be also a very good receiver, making it a much sophisticated
and heavier object. Both the system may be applied for a most stable
system and a better position evaluation.

\end{document}